\documentstyle[10pt]{article} 
\advance\hoffset by -7mm  
\renewcommand{\title}[1]{\null\vspace{25mm}

\noindent{\large{\bf #1}}\vspace{10mm}

}
\newcommand{\authors}[1]{\noindent{\large #1}\vspace{3mm}

}
\newcommand{\address}[1]{\noindent #1\vspace{5mm}

}
\renewcommand{\abstract}[1]{

\noindent
 #1}\vspace{2mm}

\begin{document}

\newcommand{\beq}{\begin{equation}}
\newcommand{\eeq}{\end{equation}}
\newcommand{\barr}{\begin{eqnarray}}
\newcommand{\earr}{\end{eqnarray}}

\newcommand{\andy}[1]{ }


\title{A DERIVATION OF THREE-DIMENSIONAL \\ 
INERTIAL TRANSFORMATIONS}
 
\authors{{\bf F. Goy}\footnote{Financial support of the Swiss 
National Science Foundation and the Swiss Academy of Engineering Sciences.}}
\address{{\em Dipartimento di Fisica,
Universit\^^ {a} di Bari  \\
Via G. Amendola, 173 \\
I-70126  Bari, Italy \\ 
E-mail:  goy@axpba1.ba.infn.it}\\
\\
June 9, 1997}
 
\abstract{The derivation of the transformations between inertial frames made 
by Mansouri 
and Sexl is generalised to three dimensions for an arbitrary direction of the 
velocity. Assuming length contraction and time dilation to have their 
relativistic values, a set of transformations kinematically 
equivalent to special relativity is obtained. The ``clock 
hypothesis'' allows the derivation to be extended to accelerated systems. A 
theory of inertial transformations maintaining absolute simultaneity is shown 
to be the only one logically consistent with accelerated movements. Algebraic 
properties of these transformations are discussed. 
}

 

\section{\normalsize{{\bf INTRODUCTION}}}
In the last two decades, there has been a revival of ether theories. There are 
at least two important reasons for this new interest in old ideas:
\begin{itemize}
\item The discovery of the microwave background radiation enables us to define 
a pri\-vi\-ledged frame in which the temperature of this radiation is isotropic.
\item The development of the parametric test theories of special relativity of 
Mansouri and Sexl \cite{mase:77a}, and that of Robertson 
\cite{robe:49a}, are based on the 
assumption that there is at least one inertial 
frame in which light behaves isotropically. It has shown, among other things,
that if the parameters of length contraction and time dilation are 
well- testable quantities, the parameter corresponding to the synchronization 
of 
distant clocks in an inertial frame is not a testable quantity \cite{vest:93a}. 
Consequently,
one obtains a set of theories equivalent to special relativity, at least in 
inertial frames. The use of the 
so-called ``absolute'' synchronization, which preserves simultaneity, leads 
to an 
ether theory in the sense that it singles out a priviledged inertial frame,
in which the one-way velocity of light is $c$ in all directions. In all other 
frames, the one-way velocity of light is generally direction-dependent; only 
the two-way velocity of light is constant in all directions and for all frames.
This theory of inertial transformations (IT) explains the 
Michelson-Morley experiment and many others (R\"{o}mer, stellar 
aberration \cite{fgoy:96a}, Fizeau, Kennedy-Thorndike, 
Ives-Stillwell, M\"{o}ssbauer rotor \cite{chis:63a},
Brillet-Hall \cite{brha:79a}, 
Hils-Hall \cite{hiha:90a},
etc...) just as well as SRT.
\end{itemize}
In their article of 1977, Mansouri and Sexl derived transformations between two 
inertial frames for an arbitrary synchronization based on two main assumptions: 
a clock moving at velocity $v$ relative to the priviledged frame slows down its 
rate according to the relativistic factor $\sqrt{1-v^{2}/c^{2}}$, and a moving 
measuring rod contracts longitudinaly with the same factor. Their 
derivation is done in the case 
in which the velocity is parallel to the $X$-axis, but they give a general 
formula 
for an arbitrary direction of the velocity without justifing it. In particular, 
it is not clear in their article, to what axes the 
coordinates are refered. The extension to an arbitrary direction of the 
velocity is not trivial, since in special relativity the product 
of two boosts with non-collinear velocity is not a boost, but it is the case 
for 
parallel velocities. Moreover, because of the Lorentz contraction it is easy to 
show that an 
orthogonal system of axes moving with constant velocity relative to an 
inertial frame appear to be non-orthogonal if the velocity of the moving 
system is not parallel to one of the axes.

In sections 2, 3 and 4 we generalise the derivation made by Mansouri and Sexl 
to an 
arbitrary direction of the velocity. In section 5, we show that only 
the inertial transformations are consistent with the clock hypothesis when 
accelerations are involved. In sections 6, analysing 
basic algebraic properties of the inertial transformations, we discuss the 
question of the 
Thomas precession in the context of the inertial theory. In section 7, 
we discuss the advantages and disavantages of the inertial theory as well as 
its domain of application.

\section{\normalsize{{\bf DERIVATION IN THE GENERAL CASE}}}

Following Robertson, and Mansouri and Sexl, we suppose that there is at least 
one privileged inertial frame $\Sigma$ in which light behaves isotropically
and in which clocks are consequently synchronized with Einstein's 
procedure \cite{eins:05a,mase:77a}. 
The length of an object having an arbitrary motion in $\Sigma$ is defined, as 
usual, as the distance measured with unit rods at rest in $\Sigma$ between the 
simultaneous (in $\Sigma$) position of the endpoints of the object.
Without loss of generality, we can choose an othonormal set ${\bf E_{1}}, 
{\bf E_{2}}, {\bf E_{3}}$  in $\Sigma$ to which the coordinates of a point are 
refered. By ``vector'' we shall mean, in the following, not a triplet of 
numbers, but a geometrical object, representing a physical one and 
independent of the system of coordinates. 
We suppose further that space is homogeneous and isotropic and that 
time is homogeneous. Another inertial system $S$ is given 
system of axes whose origin coincides with the origin 
of $\Sigma$ at time $T=0$ of $\Sigma$ and which we shall not specify further
for the time being.
If we write the coordinates of a point 
in $\Sigma$ as $(X^{\alpha}, T)$ and as $(x^{\alpha}, t)$ in $S$ 
$(\alpha = 1, 2 ,3)$, then the conditions on the origin, and of homogeneity 
and isotropy, imply 
that the relations between the cordinates are linear \cite{sell:94a}. 
We can write:
\andy{tarpet}
\barr
x^{\alpha}& =& a^{\alpha}_{\; \beta} X^{\beta} +b^{\alpha} T \nonumber\\
t\;\;& =& s_{\alpha}X^{\alpha} + a^{0}_{\;0} T\;\;\;,
\label{eq:tarpet}
\earr
where repeated indices are summed over $1,2,3$ and the co\-ef\-fi\-ci\-ents 
$a^{\alpha}_{\; \beta},\; b^{\alpha},$ $s_{\alpha}$ and $a^{0}_{\;0}$ do not 
depend on the coordinates. The matrix $a^{\alpha}_{\; \beta}$ contains 
information about longitudinal and transversal lenght contraction factors, as 
well as on the choice of the system of axes in $S$. $a^{0}_{\;0}$ is related to 
time dilation. The parameters $b^{\alpha}$ describe the velocity of the origin
of the axes of $S$ and the $s_{\alpha}$ are parameters caracterising the 
synchronization of clocks in $S$. Indices are formaly lowered and raised with 
diag(1,1,1): for example: $s^{\alpha}=\delta^{\alpha \beta}s_{\beta}$, so that 
covariant and contravariant components have numerically the same value. 

We suppose further that the origin of $S$ is moving with a velocity 
${\bf v}= v^{\alpha}{\bf E_{\alpha}}$. Imposing this condition in 
(\ref{eq:tarpet}) we obtain:
\andy{petard}
\barr
x^{\alpha}& =& a^{\alpha}_{\; \beta} (X^{\beta} -v^{\beta} T) \nonumber\\
t\;\;& =& s_{\alpha}X^{\alpha} + a^{0}_{\;0} T
\label{eq:petard}
\earr

We suppose now that {\em bodies moving through $\Sigma$ are contracted 
parallel to their motion in $\Sigma$ by the relativistic factor
$R=$$\sqrt{1-v^{2}/c^{2}}$}. In order to express it mathematically, we choose a 
set 
$({\bf e_{1},\;e_{2},\;e_{3}})$ of vectors coincident with 
$({\bf E_{1},\;E_{2},\;E_{3}})$ at $T=0$ if ${\bf v = 0}$. Because 
accelerations {\em per se} have no effects of contraction on ideal rods
\cite[p. 256]{moll:72a}, the system chosen is in 
fact the system we would obtain after acceleration up to velocity 
${\bf v}$ 
if the two systems coincided originally. We can decompose the 
vectors ${\bf E_{\alpha}}$ in a part ${\bf E_{\alpha\|}}$ parallel to ${\bf v}$
and a part ${\bf E_{\alpha \bot}}$ perpendicular to ${\bf v}$:
\andy{canne}
\beq
{\bf E_{\alpha \|}=(\hat{v} \cdot E_{\alpha}) \hat{v}}
\label{eq:canne}
\eeq
and
\andy{cone}
\beq
{\bf E_{\alpha \bot} = E_{\alpha} -(\hat{v} \cdot E_{\alpha}) \hat{v}}\;\;\;,
\label{cone}
\eeq
where ${\bf \hat{v}= v/|v|}$. So our last assumption is equivalent to:
\andy{spliff}
\barr
{\bf e_{\alpha}}& =& R{\bf E_{\alpha \|}} + {\bf E_{\alpha \bot}} \nonumber\\ 
&=& [\delta^{\beta}_{\alpha} +(R-1) \hat{v}_{\alpha} \hat{v}^{\beta}]
{\bf E_{\beta}}
\label{eq:spliff}
\earr
	\vspace{-1cm}
        \begin{figure}[ht]
        \let\picnaturalsize=N
        \def\picsize{8.0cm}
        \def\picfilename{lfig1.eps}
        \ifx\nopictures Y\else{\ifx\epsfloaded Y\else\input epsf \fi
        \let\epsfloaded=Y
        \centerline{\ifx\picnaturalsize N\epsfxsize \picsize\fi
        \epsfbox{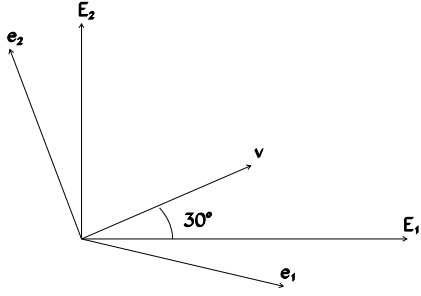}}}\fi
\caption{Configuration of the axes in two dimensions
 for $R=1/2$ and the velocity ${\bf v}$ 
making an angle of $30^{o}$ with ${\bf E_1}$.}
\end{figure}

This last equation gives us the relation between the vectors of the two 
bases. From (\ref{eq:spliff}) and (\ref{eq:petard}), 
we obtain the transformations of the 
coordinates refered to these axes, by expressing the position of a point in the 
two bases:
\andy{stick}
\barr
x^{\alpha}&=&[\delta^{\alpha}_{\beta}+(\frac{1}{R}-1)\hat{v}^{\alpha}
              \hat{v}_{\beta}]
              X^{\beta}-\frac{1}{R}v^{\alpha}T \nonumber\\
t\;\;& =& s_{\alpha}X^{\alpha} + a^{0}_{\;0} T
\label{eq:stick}
\earr
The last equation of (\ref{eq:stick}) can be rewritten as:
\andy{canna}
\beq
t = s_{\alpha}(X^{\alpha}-v^{\alpha}T) + (s_{\alpha}v^{\alpha}+ a^{0}_{\;0})T
\label{eq:canna}
\eeq

We now suppose that {\em a clock identical to the 
clocks in $\Sigma$ is at the origin of $S$ and is slowed down by a factor $R$}. 
We obtain from (\ref{eq:canna}) for the time coordinate difference that:
\andy{tulipe}
\beq
\Delta t = (s_{\alpha}v^{\alpha}+ a^{0}_{\;0})\Delta T = R \Delta T\;\;\;,
\label{eq:tulipe}
\eeq
so that (\ref{eq:stick}) can now be written as:
\andy{moustache}
\barr
x^{\alpha}&=&[\delta^{\alpha}_{\beta}+(\frac{1}{R}-1)
              \hat{v}^{\alpha}\hat{v}_{\beta}]
              X^{\beta}-\frac{1}{R}v^{\alpha}T \nonumber\\
t\;\;& =& s_{\alpha}(X^{\alpha}-v^{\alpha}T)+R T
\label{eq:moustache}
\earr
Only the tree parameters $s_{\alpha}$ caracterising the synchronization in $S$
remain unknown. The inverse of (\ref{eq:moustache}) is given by:
\andy{shilom}
\barr
X^{\alpha}&=&[\delta^{\alpha}_{\beta}+(R-{\bf s \cdot v})(1-\frac{1}{R})
            \hat{v}^{\alpha}\hat{v}_{\beta}-v^{\alpha}s_{\beta}]x^{\beta}
            +\frac{v^{\alpha}}{R}t \nonumber\\
T\;\;&=&[(\frac{1}{R}-1)\hat{v}_{\alpha}({\bf s \cdot \hat{v}})-
\frac{s_{\alpha}}{R}
]x^{\alpha}+\frac{1}{R}\;\;t\;\;\;,
\label{eq:shilom}
\earr
where ${\bf s} = s^{\alpha}{\bf E_{\alpha}}$.
In (\ref{eq:moustache}) and (\ref{eq:shilom}), the inertial 
transformations are given by ${\bf s = 0}$ , while the Lorentz transformations 
are given by ${\bf s = -v}/c^{2}R$.

\section{\normalsize{{\bf ONE-WAY VELOCITY OF LIGHT IN S}}}

Given the isotropy of the velocity of light in $\Sigma$, two events 
which are light-like obey the equation:
\andy{sipsi}
\beq
\Delta X^{\alpha}\Delta X_{\alpha}-c^{2}\Delta T^{2} = 0\;\;,
\label{eq:sipsi}
\eeq
where $\Delta$ expresses the difference between the coordinates of the two 
events. 
In order to calculate the left-hand-side of (\ref{eq:sipsi}) in $S$, we define 
new parameters $h_{\alpha}$ caracterising the synchronization in $S$, which are 
more convienient than the $s_{\alpha}$ for this calculation \cite{anst:92a}:
\andy{afghan}
\beq
{\bf h}= -(cR {\bf s}+\frac{{\bf v}}{c})
\label{eq:afghan}
\eeq
The Lorentz transformations are obtained for ${\bf h = 0}$, while the inertial 
transformations are obtained for ${\bf h}=-{\bf v}/c$. If we have 
synchronized the clocks with ${\bf h = 0}$ in $S$, we can obtain an 
``absolute'' 
synchronization ${\bf h}=-{\bf v}/c$ by performing a resynchronization of 
our clocks.
For a general ${\bf h}$ the resynchronization is given by:
\andy{ktama}
\beq
ct^{{\bf h}} = ct^{{\bf h = 0}} - {\bf h \cdot x}
\label{eq:ktama}
\eeq
For the calculation of (\ref{eq:sipsi}), we can first use a Lorentz boost to 
reach $S$. We obtain:
\andy{narguille}
\beq
\Delta x^{\alpha}\Delta x_{\alpha}-c^{2}\Delta (t^{{\bf h=0}})^{2} = 0\;\;\;,
\label{eq:narguille}
\eeq
because the Lorentz transformations leaves the Minkowski metric invariant. 
Resynchronizing the clocks in $S$ with (\ref{eq:ktama}), we obtain the equation
:
\andy{libanais}
\beq
c^{2}\Delta t^{2} +2c \Delta t {\bf h \cdot \Delta x} + 
({\bf h \cdot \Delta x})^{2}- ({\bf \Delta x})^{2} = 0\;\;\;,
\label{eq:libanais}
\eeq
which can be written with the metric:
\andy{junk}
\barr
g_{\alpha \beta}= \left( \begin{array}{cccc}
-1      & -h_{1}        & -h_{2}         & -h_{3} \\
-h_{1}  & 1-h_{1}^{2}   & -h_{1}h_{2}    & -h_{1} h_{3} \\
-h_{2}  & -h_{1}h_{2}   & 1-h_{2}^{2}    & -h_{2} h_{3}  \\
-h_{3}  & -h_{1}h_{3}   & -h_{2} h_{3}   & 1-h_{3}^{2} 
\end{array} \right)
\label{eq:junk}
\earr

From (\ref{eq:libanais}), the velocity of light in $S$ can easily be calculated
and is given in the direction ${\bf \hat{n}} = \Delta x^{\alpha}{\bf 
e_{\alpha}}/\|{\bf x}\|$ by:
\andy{marocain}
\beq
{\bf c(\hat{n}}) =\frac{ c {\bf \hat{n}}}{1-{\bf h \cdot \hat{n}}}
\label{eq:marocain}
\eeq
First, note that the one-way velocity of light is generally direction 
dependent in 
$S$. The two-way velocity of light is on the other hand constant in all 
directions.
The inverse of the two-way velocity of light 
${\bf \stackrel{\leftrightarrow}{c}( \hat{n})}$ in direction ${\bf \hat{n}}$ is 
given by:
\andy{skunk}
\beq
\frac{1}{\|{\bf \stackrel{\leftrightarrow}{c}( \hat{n})}\|} =
 \frac{1}{2}\left(\frac{1}{\|{\bf c(\hat{n})}\|}+
\frac{1}{\|{\bf c(-\hat{n})}\|}\right)
=\frac{1}{c}\;\;\;,
\label{eq:skunk}
\eeq
in all directions. So we see that assuming the length contraction and time 
dilation parameters to have their relativistic value implies a two-way velocity 
of light constant in all directions. Other derivations had already shown that,
assuming the time dilation parameter to have its relativistic value, and 
assuming also the constancy of the two-way velocity of light, the 
length contraction parameters have their relativistic values \cite{sell:95a}.

Secondly, notice that the norm of a vector ${\bf x}$ in $S$, whose 
coordinates are $x^{\alpha}\;(\alpha=1,2,3)$ is given by 
$\|{\bf x}\|=\sqrt{x^{\alpha}x_{\alpha}}$ despite the fact that, seen from 
$\Sigma$, the set ${\bf e_{\alpha}}\;(\alpha=1,2,3)$ is not orthonormal. But 
in $S$, lengths are measured with unit rods at rest in $S$. These measuring 
instruments as well as the objects they measure are contracted relative to 
$\Sigma$, so that the set ${\bf e_{\alpha}}\;(\alpha=1,2,3)$ is found to be 
orthonormal in $S$ and consequently the norm of a vector is calculated with a 
diagonal 3-dimensional metric. A mathematical proof of this last remark 
can be given making use of (\ref{eq:junk}). As is well known, in the 
formalism of general relativity, which includes transformations 
(\ref{eq:moustache}) as a special case, the spatial part $\gamma_{ij}$
of the metric of a four 
dimensional space-time metric is not only given by the space-space coefficients 
of this metric but by the expression:
\andy{freak}
\beq
\gamma_{ij}=\left(g_{ij}-\frac{g_{i0}g_{j0}}{g_{00}}\right)\;\;
i,\;j = 1,\;2,\;3\;\;\;,
\label{eq:freak}
\eeq
which gives, in the case of (\ref{eq:junk}), diag(1,1,1); that is, 
an orthonormal 
metric in Cartesian coordinates. Eq. (\ref{eq:freak}) corresponds to the fact 
that distances are measured with rigid measuring rods 
(see \cite[p. 269]{moll:72a}), 
which can be defined in terms of two-way light signals \cite{anst:92a}.

Note also that in (\ref{eq:junk}), the components of the vector potential (or 
gravitomagnetic potential) $(g_{01},g_{02},g_{03})$ are non zero. The 
gravitomagnetic field \cite{ciwe:95a} is defined as ${\bf H} = {\bf \nabla} 
\wedge (-{\bf h})$. As in the case of the Minkowski metric, we have from 
(\ref{eq:junk}): ${\bf H= 0}$, whenever $(-{\bf h})$ can be 
written in the form $-{\bf h}={\bf \nabla}f$, where f is a function.

\section{\normalsize{{\bf SYNCHRONISATION WITH ONE PARAMETER}}}

If we now assume that {\em there is in $S$ only one privileged direction given
by the velocity ${\bf v}$}, we can restrict the three parameters $s_{\alpha}$
to one parameter. We suppose that the one-way velocity of light has the same 
norm for any direction making the same angle with ${\bf v}$. This means that
${\bf \hat{n}_{1} \cdot h = \hat{n}_{2}\cdot h}$ 
$\forall \; {\bf \hat{n}_{1},\; \hat{n}_{2}},$ such that ${\bf \hat{n}_{1}\cdot
v=\hat{n}_{2}\cdot v}$. For ${\bf h=0}$, this condition is trivially fulfilled, 
otherwise it is equivalent to ${\bf h} = h {\bf \hat{v}}$, so that also 
${\bf s} = s {\bf \hat{v}}$.
We remain with only one parameter characterising the synchronization and 
equation 
(\ref{eq:moustache}) becomes:
\andy{kif}
\barr
x^{\alpha}&=&[\delta^{\alpha}_{\beta}+(\frac{1}{R}-1)
              \hat{v}^{\alpha}\hat{v}_{\beta}]
              X^{\beta}-\frac{1}{R}v^{\alpha}T \nonumber\\
t\;\;& =& s \hat{v}_{\alpha}X^{\alpha}+(R-vs) T
\label{eq:kif}
\earr
with its inverse given by:
\andy{beuh}
\barr
X^{\alpha}&=&[\delta^{\alpha}_{\beta}+(R-vs-1)
            \hat{v}^{\alpha}\hat{v}_{\beta}]x^{\beta}
            +\frac{v^{\alpha}}{R}t \nonumber\\
T\;\;&=&-s(\hat{v}_{\alpha}x^{\alpha})+\frac{1}{R}\;t
\label{eq:beuh}
\earr
Substituing $s=-v/c^{2}R$ in (\ref{eq:kif}), we obtain a Lorentz-boost in (3+1) 
dimension (see \cite[eqs:(2.1.20),(2.1.21)]{wein:72a}):
\begin{eqnarray}
x^{\alpha}&=&[\delta^{\alpha}_{\beta}+(\frac{1}{R}-1)
              \hat{v}^{\alpha}\hat{v}_{\beta}]
              X^{\beta}-\frac{1}{R}v^{\alpha}T \nonumber\\
t\;\;& =& \frac{1}{R}\left(T-\frac{v_{\alpha}X^{\alpha}}{c^{2}}\right)
\label{eq:kof}
\end{eqnarray}
Substituing $s=0$, we obtain the (3+1) dimensional inertial transformation:
\begin{eqnarray}
x^{\alpha}&=&[\delta^{\alpha}_{\beta}+(\frac{1}{R}-1)
              \hat{v}^{\alpha}\hat{v}_{\beta}]
              X^{\beta}-\frac{1}{R}v^{\alpha}T \nonumber\\
t\;\;& =& R\; T
\label{eq:kaf}
\end{eqnarray}
In the following table, we show also for which value of $h$ the Lorentz 
transformation and the inertial transformation are recovered.
\begin{center}
\begin{tabular}{|c||c|c|}                            \hline
         & LT                  & IT             \\   \hline \hline
$h$      & $0$                 & $-\frac{v}{c}$ \\   \hline
$s$      & $-\frac{v}{c^{2}R}$ & $0$            \\   \hline
\end{tabular}
\end{center}
Different choices of $s$ represent different conventions of synchronization of 
distant clocks, corresponding to different values of the one-way speed of 
light. In theories in which no interaction between the measuring apparatus and 
the system is assumed, the setting of clocks cannot change the objective nature 
of physical processes, but only our description of it. Thus, the transformations
(\ref{eq:kof}) and (\ref{eq:kaf}) represent the same transformation of 
spacetime, but relative to different spacetime coordinates resulting from the 
choice of different methods of synchronization of clocks. The above remark is 
not restricted to the two particular values of $s$ in equations (\ref{eq:kof}) 
and (\ref{eq:kaf}) and we can say that (\ref{eq:kif}) represents a set of 
transformations equivalent to the transformation of Lorentz  between $\Sigma$ 
and $S$. Strictly speaking, this assertion is true in inertial systems only, 
since our derivation of (\ref{eq:kif}) was made in this class of systems.
Different coordinatisations than that of the Lorentz transformation are 
obviously equivalent: nobody claims, for example, that we obtain different 
physical
results if we use three axes which are not orthonormal. The intrasystemic
synchronization of distant clocks, by defining the simultaneity of distant 
events, allows the attribution of a common time to all points of an inertial 
system and thus corresponds to the choice of an axis in the spacetime domain.
Without
synchronizing clocks time would not even be defined, but we would only have
a ``time'' for each clock with no connection between them and thus no way of 
attributing a time coordinate to an event. In relativity and 
related theories, clocks are synchronized by means of light signals whose 
one-way velocity is {\em assumed} as stressed by Einstein in 1916 
\cite{eins:16a}:
{\em ``that light requires the same time to traverse the path AM...as the path 
BM [M being the midpoint of the line AB] is in reality neither a supposition 
nor 
a hypothesis about the physical nature of light, but a stipulation which I can 
make of my own free will''.} Numerous authors have stressed the conventional
nature of the one-way speed of light (for a review see: \cite{sell:96a}).
The explaination of Anderson and Stedman is of particular interest 
\cite{anst:92a}: {\em `` In order to synchronize spatially separated clocks, 
signals must be sent between them and the time of passage of such signals must 
be known. A maximum specification will only arise through the use of light 
signals by virtue of their property of being first signals. But we are caught 
in a circle here as the time of passage for first signals can only be obtained 
by prior knowledge of the synchronization of clocks. The essence of the 
``conventionalist'' position is that this circularity is inescapable and that 
no fact of nature permits a unique determination of either the simultaneity  
relation within an inertial frame or the speed of light in a given direction''}.
So the freedom of choice of the synchronization of clocks is equivalent to the 
freedom of choice in the assumption of the value of the one-way speed of light,
which leads to a freedom in the choice of the spacetime coordinate system. From 
this point of view, it is obvious that (\ref{eq:kif}) represents a set of 
transformations experimentally
equivalent to the Lorentz transformations between $\Sigma$ and 
$S$.

\section{\normalsize{{\bf THE CLOCK HYPOTHESIS}}}

In this section, we will show that in the case of accelerated systems, 
only the transformation with $s=0$ is consitent with the ``clock hypothesis''.
In this sense of logical consistency, transformations (\ref{eq:kof}) and 
(\ref{eq:kaf}) are no more equivalent when extended to accelerated systems.

The ``clock hypothesis'' states that the rate of an ideal clock accelerated 
relative to an inertial frame is identical to the rate of a similar clock in 
the instantaneously comoving inertial frame. With other words, the rate of 
clocks is not influenced by accelerations {\em per se}, when seen from inertial 
frames. It also supposed that real clocks exist in nature, which approach
the conditions of the clock hypothesis. To our knowledge, this assumption was
first implicitly used by Einstein in 1905 \cite{eins:05a} and was superbly 
confirmed in the 
CERN muon storage ring experiment \cite{bail:77a}, where the muons had a 
time of decay depending only on their 
velocity (in agreement with the time dilation formula) despite the fact that 
their acceleration was of $10^{18}g$. We stress here that the clock hypothesis 
is an assumption logically independent of the assumptions used to derive
(\ref{eq:kif}), and of the two postulates of special relativity.

Using the clock hypothesis in order to extend the theory to the possibility 
of temporary acceleration, we will restrict the last unknown parameter. The 
necessity of this extension is obvious, since a frame needs to accelerate to go 
from an uniform velocity to an other uniform velocity. Developing an argument 
of Bell \cite{bell:88a} and Selleri \cite{sell:95a}, we imagine the following 
situation. Two clocks are in 
two spaceships at rest in $\Sigma$. The first clock A is at the origin of 
$\Sigma$
and the second clock B at a point P, whose position is given by a vector 
${\bf d}$. At $T=0$, the two spaceships begin to accelerate in straight line 
under the constraint of a constant force equal for both spaceships. They 
continue their acceleration up to a time $\bar{T}$, and then continue in 
uniform 
motion. In $\Sigma$ the law of motion of relativity is valid because clocks are 
synchronized with an Einstein's procedure. It follows that the two spaceships 
will have an hyperbolic motion whose equation is given for the time 
$T\le\bar{T}$
by:
\andy{pete}
\barr
{\bf X}_{A}(T)&=&\frac{{\bf \hat{n}}c^{2}}{a}\{[1+(\frac{aT}{c})^{2}]
^{\frac{1}{2}}-1\} \nonumber \\
{\bf X}_{B}(T)&=&\frac{{\bf \hat{n}}c^{2}}{a}\{[1+(\frac{aT}{c})^{2}]
^{\frac{1}{2}}-1\} +{\bf d}\;\;\;,
\label{eq:pete}
\earr
where ${\bf \hat{n}}$ is a unit vector caracterising the direction of 
propagation and $a$ is the ratio of the force by the rest mass of the 
spaceship. For $T>\bar{T}$, the same term ${\bf \hat{n}}v(T-\bar{T})$ ($v$ 
is the velocity attained at $T= \bar{T}$) must be 
added to both equations (\ref{eq:pete}) so that
$\forall T:\; {\bf X}_{B}(T)-{\bf X}_{A}(T)=
{\bf X}_{B}(0)-{\bf X}_{A}(0)={\bf d}$. That is the distance between the 
spaceships remains constant in $\Sigma$.
The velocity of the spaceships for $0 \le T \le \bar{T}$ is given by:
\andy{nase}
\beq
{\bf \dot{X}}_{A,B}(T) =\frac{{\bf \hat{n}}aT}{[1+(\frac{aT}{c})^{2}]
^{\frac{1}{2}}}
\label{eq:nase}
\eeq
The clock hypothesis states mathematically that:
\andy{fait}
\beq
dt =\left(1-\frac{\dot{X}_{A,B}^{2}(T)}{c^{2}}\right)^{\frac{1}{2}}dT\;\;\;,
\label{eq:fait}
\eeq
where $t$ is the time coordinate in the accelerated system S.
Integrating (\ref{eq:fait}) with (\ref{eq:nase}) and the given initial 
conditions, we obtain for the time of $S$ as function of the time of $\Sigma$, 
for $0\le T \le \bar{T}$:
\andy{loaded}
\beq
t_{A}(T)=t_{B}(T)=\frac{c}{a}\;arcsinh(\frac{aT}{c})
\label{eq:loaded}
\eeq
and for $T>\bar{T}$:
\andy{pollen}
\beq
t_{A}(T)=t_{B}(T)=\frac{c}{a}\;arcsinh(\frac{a\bar{T}}{c}) + 
\sqrt{1-v^{2}/c^{2}}\;\;(T-\bar{T})
\label{eq:pollen}
\eeq
so that $\forall T:\; t_{A}(T) - t_{B}(T)=0$.
Thus we see that though slowing their rate, clocks A and B maintain the 
simultaneity of events.

On the other hand, we can calculate the difference of time coordinates in $S$ 
using directly an inertial transformation with a general $s$.
Using (\ref{eq:kif}), we find for two simultaneous events in $\Sigma$
\andy{stone}
\beq
t^{s}_{B}(T)-t^{s}_{A}(T)=s {\bf \hat{v}} \cdot ({\bf X}_{B}(T)-{\bf X}_{A}(T))
=s {\bf \hat{v} \cdot d}
\label{eq:stone}
\eeq
Since this last equation must hold for all ${\bf d}$, we obtain $s=0$ 
comparing with (\ref{eq:loaded}).
In particular, an observer moving with the spaceships would have to 
resynchronize the clocks in order to obtain the Lorentz-transformations 
between his frame and $\Sigma$ \cite{mast:93a}. Nevertheless, such a 
resynchronization allows to define a time 
coordinate only locally as explained by Misner, Thorne and Wheeler 
\cite{mith:73a}:
{\em ``Difficulties also occur when one considers an observer who begins at 
rest 
in one frame, is accelerated for a time, and maintains thereafter a constant
velocity, at rest in some other inertial coordinate system. Do his motion 
define in any natural way a coordinate system? Then this coordinate system (1) 
should be the inertial frame $x^{\mu}$ in which he was at rest for time $x^{0}$
less than 0, and (2) should be the other inertial frame $x^{\mu'}$ for times
$x^{0'}>T'$ in which he was at rest in the other frame. Evidently some further 
thinking would be required to decide how to define the coordinates in the 
regions not determined by these two conditions. More serious, however, is the 
fact that these two conditions are inconsistent for a region of spacetime that 
satisfies simultaneously $x^{0}<0$ and $x^{0'}>T'$.''} Further they conclude 
that
the problem possess a solution only in the immediate vicinity of the observer.
With other words special relativity is valid only locally when accelerations 
come into
play. That is indeed one of the reasons why special relativity is valid only 
locally within the frame of general relativity. The reader  
thinks perhaps that the problem above does not occur in general relativity, 
where there is absolutely no 
doubt that accelerated observers can be considered. 
General relativity is
coordinate independent and thus any synchronization can be used. The latter 
statment is true only 
locally, because some synchronizations allow a global definition of time, 
others (in particular Einstein's synchronization) are path dependent when the 
metric is non-static \cite{clas:00a}\cite[p. 379]{moll:72a}. This implies in 
particular that the binary relation between two events
``A is synchronous with B'' is not an equivalence relation in the latter case.
Applying the formalism of general relativity to 
the rotating disk, one can show that
only a synchronization corresponding to $s=0$ allows a global 
and self-consistent definition of time on the disk \cite{fgoy:97a}. 
Synchronization problems in the frame of general relativity is beyond the scope 
of this paper. We have mentioned them shortly in order to show that the problem
exponed in this section does not disappear in general relativity. 

Coming back to the problem of two accelerated clocks treated in this section we 
conclude 
that the only theory whose 
synchronization of clocks is {\em logically consistent} with the clock 
hypothesis is the inertial theory using $s=0$. With other word, the natural 
behaviour of clocks (that is without resynchronisation by an observer) 
is described by a theory using an ``absolute'' simultaneity ($s=0$). Moreover a 
time defined in this way is global as it should be. 

This result could have been derived with weaker assumptions than the clock 
hypothesis. For example, we would have obtained the same result if the rate 
of 
the clocks were dependent on their accelerations (seen from an inertial frame).
What is 
important in the above reasoning is that clocks A and B, which are 
Einstein's synchronised in 
$\Sigma$, are submitted to the 
same influences, so they have the same rate, so maintain the simultaneity of 
$\Sigma$. This idea was already expressed by Einstein in his article of 1907
where he first used the principle of equivalence and discussed the behaviour 
of two clocks in an accelerated frame $\Sigma$ which was at rest in an inertial 
frame $S$ at time $t=0$ (the clocks are Einstein's synchronised in $S$). 
He wrote \cite{eins:07a} {\em ``Indeed, being given 
that two arbitrary clocks of $\Sigma$ are synchronous at time $t=0$ 
with respect to $S$ and 
that they are subject to the same motions, they remain constantly synchronous
with respect to S. But for this reason, and according to the paragraph 4,} 
[that is according to the Lorentz transformations] {\em they do 
not work in a synchronous way with respect to a system $S'$ momentarily at rest 
with respect to $\Sigma$ but in motion with respect to $S$.''}

\section{\normalsize{
{\bf SOME ALGEBRAIC PROPERTIES OF INERTIAL\\
 TRANSFORMATIONS}}}

We call $A({\bf v})$ the inertial transformation (\ref{eq:kif}) between 
$\Sigma$ and $S$:
The transformations $\Omega({\bf v},{\bf w})$ between two inertial systems $S$ 
and $S'$ having a velocity 
${\bf v}$ and ${\bf w}$ relative to $\Sigma$ can be found by elimination of the 
coordinates of the $\Sigma$ system as:
\andy{ocb}
\beq
\Omega({\bf v},{\bf w})= A({\bf v})A^{-1}({\bf w})
\label{eq:ocb}
\eeq
and have in general a complicated form. Some of the algebraic properties of the 
set of all transformations $\Omega({\bf v},{\bf w})$ can be demonstrated using 
the definition (\ref{eq:ocb}). Notice that each element 
$\Omega({\bf v},{\bf w})$ is parametrized by 6 parameters $v^{\alpha},
w^{\alpha}$, which can vary continously, subject only to the conditions
$\|{\bf v}\|\leq c$ and $\|{\bf w}\|\leq c$.

It does not make physically sense to multiply $\Omega({\bf v},{\bf w})$ :
$S \rightarrow S'$ with $\Omega({\bf u},{\bf r})$ : $S'' \rightarrow S'''$.
Mathematically, we can multiply two such transformations. Using (\ref{eq:ocb}),
we obtain:
\andy{fume}
\beq
\Omega({\bf v},{\bf w})\Omega({\bf u},{\bf r})= A({\bf v}) A^{-1}({\bf w})
A({\bf u})A^{-1}({\bf r})\;\;\;,
\label{eq:fume}
\eeq
which is in general not a transformation of type $\Omega$, but an object 
depending on 12 independant parameters.
So the set of all transformations $\Omega$, 
with the composition law defined above, {\em does not form a group}
because it is not closed.
What makes physically sense is to go from $S \rightarrow S'$ and then from
$S' \rightarrow S''$. In this case, the composition of two 
transformations $\Omega({\bf v},{\bf w})$: $S \rightarrow S'$ and
$\Omega({\bf w},{\bf u})$: $S' \rightarrow S''$:
\andy{rizla}
\beq
\Omega({\bf v},{\bf w})\Omega({\bf w},{\bf u})= A({\bf v}) A^{-1}({\bf w})
A({\bf w})A^{-1}({\bf u})=\Omega({\bf v},{\bf u})\;\;\;,
\label{eq:rizla}
\eeq
gives again a transformation of type $\Omega$:
$\Omega({\bf v},{\bf u})$ from $S$ to $S''$.
The properties of the inertial transformations are 
similar to those of evolution operators.
Despite of the fact that the $\Omega$'s do not form a group, for the cases 
which are physically relevant, everything is well defined.
We have neutral 
elements, every element has an inverse and the composition is associative:
\begin{enumerate}
\item $\Omega({\bf v},{\bf v})=I\;\; \forall {\bf v}$

\item $\Omega({\bf v},{\bf w})\Omega({\bf w},{\bf v}) =I\;\;\forall {\bf v},
{\bf w}$
 
\item 
$[\Omega({\bf v},{\bf w})\Omega({\bf w},{\bf u})]
\Omega({\bf u},{\bf r})
=\Omega({\bf v},{\bf w})[\Omega({\bf w},{\bf u})\Omega({\bf u},{\bf
r})]\\
=\Omega({\bf v},{\bf r}) \;\;\;   \forall {\bf v},\;{\bf w},\;{\bf u},\;
{\bf r}$
\end{enumerate}
Moreover we have det$\Omega({\bf v},{\bf w})=1$, wheras the determinant of the 
spatial part only is given by $R(v)/R(w)$. Note that a transformation 
$A({\bf v})$: $\Sigma \rightarrow S$ can also be expressed with the following
formula:
\andy{smoking}
\beq
R_{{\bf \hat{n}}}(\Psi)A(R_{{\bf \hat{n}}}(-\Psi){\bf v})R_{{\bf \hat{n}}}
(-\Psi)=A({\bf v}),
\label{eq:smoking}
\eeq
where ${\bf \hat{n}}$ is a unit vector and $R_{{\bf \hat{n}}}(\Psi)$ is 
a rotation around this vector of an angle $0 \leq \Psi < \pi$, measured in 
$\Sigma$. As a special case of equation (\ref{eq:smoking}), we can express
$A({\bf v})$ as given by (\ref{eq:kif}) in terms of a inertial boost 
along the $X$-axis and a rotation.

So far we have always considered axes of $\Sigma$ and $S$ in 
the positions described in section 2: they would coincide if their 
relative velocity vanishes. If we want to describe  transformations of 
coordinates in which the axes ${\bf e}_{\alpha}$ are rotated to new axes
${\bf e'}_{\alpha}$ such that ${\bf e'}_{\alpha}=R_{{\bf \hat{n}}}(\Psi)
{\bf e}_{\alpha}$, then this more general transformation $K$ from 
$\Sigma$ to $S$ is 
the combination of a ``boost'' and a rotation. We have:
\andy{taf}
\beq
K({\bf v},{\bf \hat{n}},\psi) =   R_{{\bf \hat{n}}}(-\Psi)A({\bf v})
=A(R_{{\bf \hat{n}}}(-\Psi){\bf v}) R_{{\bf \hat{n}}}(-\Psi)
\label{eq:taf}
\eeq
From (\ref{eq:taf}) the corresponding type of transformations between $S$ and 
$S'$, J for 
axes with arbitrary rotations, will be of the type:
\andy{bogart}
\beq
J=K K'^{-1}= A({\bf v}) R_{{\bf \hat{m}}}(\Theta)A^{-1}({\bf w})
\label{eq:bogart}
\eeq

We have seen in (\ref{eq:rizla}) that the composition of two transformations 
without rotation,
$\Omega({\bf v},{\bf w})$: $S \rightarrow S'$ and $\Omega({\bf w},{\bf u})$: 
$S' \rightarrow S''$ gives a transformation $\Omega({\bf v},{\bf u})$: 
$S \rightarrow S''$, without rotation. This law is valid for any velocities 
${\bf v}$
,${\bf w}$, ${\bf u}$, that is also when the velocity from $S'$ relative to $S$
and the velocity of $S''$ relative to $S'$ are not collinear. This is a crucial 
difference with special relativity, where the Lorentz boosts do not form a 
subgroup of the orthochrone homogeneous proper Lorentz group \cite{unga:96a}. 
In particular, the 
composition of two Lorentz boosts with non-collinear velocities is not a Lorentz
boost, but a Lorentz boost multiplied by a rotation (A). 
This give rise to Thomas precession (B)\cite{thom:26a}. Does it mean that 
the Thomas 
precession can not be deduced from the theory of inertial transformations? No 
because in logic: (A $\Rightarrow$ B) does not imply ($\neg$A $\Rightarrow$
$\neg$B). From the absence of rotation in the multiplication of two 
``inertial boosts'' ($\neg$A), one cannot conclude to the absence of Thomas 
precession ($\neg$B).
Other authors than Thomas 
(\cite{bami:59a,schw:74a,mull:92a}, see also: \cite[ch. 11]{jack:82a} and 
\cite{fish:72a}) have given other 
derivations of the Thomas precession. In their case dynamical effects are 
explained with dynamical causes and the algebraic properties of the Lorentz 
transformations do not come directly into play. It could be 
the subject of a future article to derive the Thomas precession in the frame of 
the inertial theory. Note that the Thomas precession is in fact well tested in 
the case of the precession of the muons in the muon storage ring
(\cite{bail:79a,bami:59a}). 

\section{\normalsize{{\bf DISCUSSION}}}

In 1977, Mansouri and Sexl wrote :{\em ``A convention about clock 
synchronization can be choosen that does maintain absolute simultaneity. Based 
on this convention an ether theory can be constructed that is, as far as 
kinematic is concerned (dynamics will be studied in a later paper in this 
series), 
equivalent to special relativity''}. Note that this paper was never written, 
but the Maxwell's equations were studied by Chang \cite{chan:79a} and 
Rembieli\'nski \cite{remb:80a}, the
dynamics by the latter and Selleri \cite{sell:96b}. One can say that from a 
dynamical point of view this theory is also equivalent to special relativity.
The reader is surely asking himself what is the advantage of introducing an 
ether (priviledged frame $\Sigma$) in a theory equivalent to special 
relativity. Of course this can be done theoretically, but since this theory is 
equivalent to special relativity it ensures us also that this ether or the
velocity of an inertial frame $S$ respective to $\Sigma$ is not 
detectable within the frame of this theory. Is the development of such theories 
not a step back in the history of science? I would say that it is a step back
when we consider inertial frames only, which allow us to make two steps forward
when we consider accelerated frames. We have seen in section 5, that two clocks 
which are synchronized with an Einstein's procedure in $\Sigma$ maintain an 
absolute simultaneity respective to $\Sigma$ after and during acceleration 
and thus the theory of inertial 
transformation is advantaged in the sense that it corresponds to the behaviour 
of real clocks. This advantage is of logical nature in the sense that the 
theory of inertial transformation allows a global and self-consistent 
definition 
of time, where special relativity cannot do it. This advantage is still 
present in general 
relativity where the Einstein's synchronization does not allow to define time 
globally and self-consistently when the metric is non-static. The difference 
between the two theories is not really of experimental nature in the sense
that we cannot mesure the absolute velocity of an inertial frame $S$ with 
respect to $\Sigma$, when $\Sigma$ is interpreted as an objectively 
priviledged frame.
As already stated by Vetharaniam and
Stedman \cite{vest:93a}, the synchronization of clocks is 
also conventional in $\Sigma$, so that we can call $\Sigma$ any inertial frame:
for any  practical purposes 
$\Sigma$ is the 
inertial frame where the clocks are first synchronized with an Einstein's 
procedure. 

\section{\normalsize{{\bf CONCLUSION}}}

A derivation of the transformations between inertial 
systems was made in the general three-dimensional case. Under the main 
assumptions that there is at least one inertial frame where light behaves 
isotropically and that time dilation 
and lenght contraction have their relativistic values, we find the whole set of 
theories kinematically equivalent to special relativity. In the derivation, the 
notion of vector as an object independent of the coordinates system is 
used. This approach allows one to see clearly the sort of system of axes 
to which the 
coordinates are refered. Moreover, accelerated movements are not {\em a priori} 
exclued in this reasoning. The constancy of the two-way velocity of light is 
shown to be a consequence of the given assumptions. For a general parameter of 
synchronization, the one-way velocity of light is generally direction dependant 
and is only equal to c in the case of special relativity theory.
An extension of the transformations to accelerated 
systems, through the clock hypothesis, shows that only a theory maintaining 
absolute simultaneity is logically self-consistent.

\section{\normalsize{{\bf ACKNOWLEDGEMENT}}}
I want to thanks the Physics Departement of Bari Uni\-ver\-si\-ty for 
hospitality, 
and Prof. F. Selleri for its kind suggestions and criticisms.

\end{document}